\documentclass{egpubl}
\usepackage{amsmath}

\JournalSubmission    
 \electronicVersion 

\ifpdf \usepackage[pdftex]{graphicx} \pdfcompresslevel=9
\else \usepackage[dvips]{graphicx} \fi

\PrintedOrElectronic

\usepackage{t1enc,dfadobe}

\usepackage{egweblnk}
\usepackage{cite}

\let\it=\itshape

\newcommand{\etal}{{\it{et~al. }}}

\newcommand{\secname}{Section}
\newcommand{\figname}{Figure}
\newcommand{\tabname}{Table}
\newcommand{\eqname}{Eq.}

\newcommand{\argmin}{\mathop{\textrm{argmin}}\limits}

\usepackage{subfigure}

\graphicspath{{./figures/}}

\title[Texturing and Deforming Meshes with Casual Images]%
      {Texturing and Deforming Meshes with Casual Images}

\author[I-Chao Shen \& Yi-Hau Wang \& Yu-Mei Chen \& Bing-Yu Chen]
{\parbox{\textwidth}{\centering I-Chao Shen~
        Yi-Hau Wang ~ Yu-Mei Chen ~ Bing-Yu Chen
        }
        \\
{\parbox{\textwidth}{\centering National Taiwan University
       }
}
}

\begin{document}
\maketitle

\begin{abstract}
Using (casual) images to texture 3D models is a common way to create realistic 3D models, which is a very important task in computer graphics. However, if the shape of the casual image does not look like the target model or the target mapping area, the textured model will become strange since the image will be distorted very much. In this paper, we present a novel texturing and deforming approach for mapping the pattern and shape of a casual image to a 3D model at the same time 
based on an alternating least-square approach. 
Through a
photogrammetric method, we project the target model onto the source image according to the estimated 
camera model. Then, the target model is deformed according to the shape of the source image using a surface-based deformation method while minimizing the image distortion simultaneously. The processes are performed iteratively until convergence.
Hence, our method can achieve texture mapping, shape deformation, and detail-preserving at once, and can obtain more reasonable texture mapped results than traditional methods. 

\end{abstract} 
\section{Introduction}
Texture mapping
is a very common way to create realistic 3D models 
and
is a very important task in computer graphics.
Traditionally, the texture images should be designed carefully by artists to achieve high quality texture patterns and less distortion while mapping to a specific 3D model.
However,
in recent years, 
due to
the wide spread of digital cameras, 
the number of (casual) images on the Internet is increasing in a very high speed. 
Hence, people are thinking to map
these images 
to some template models to create various realistic 3D models.
This may lead to increase the freedom of texture mapping, 
but may also introduce serious distortion and alignment problems since such casual images are not designed well for mapping to the template models.

To obtain visually pleasing textured models with casual images, 
the serious 
distortion and alignment problems 
should be solved.
Although
many previous papers \cite{Levy:2002,Desbrun:2002,Zhang:2005} dealt with the distortion problem, 
those
methods all strived to minimize the distortion mathematically 
for mapping a texture image onto the surface of a given model.
However, if the given model does not look like the shape of the texture image, although the distortion could be minimized mathematically, the textured result will be still very strange.
This problem may become much more serious
when we 
want to use the
casual image(s) to 
texturing an arbitrary model.

\begin{figure*}[tb]
\centering
\includegraphics[width=\linewidth]{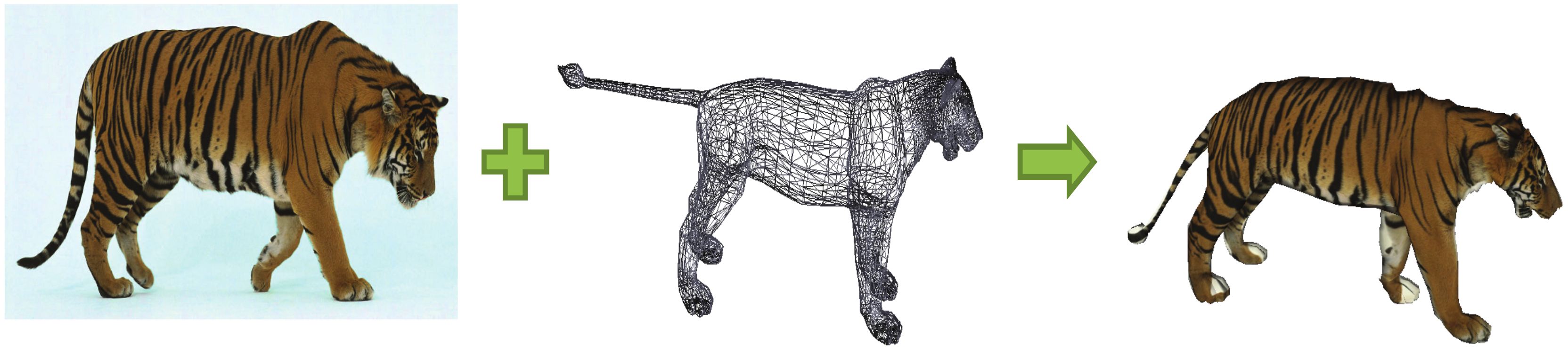}
\caption{Overview of our method. Given a casual image obtained from the Internet, and a triangle mesh model, we compute the texture mapping and shape deformation interleaving.
This example uses a tiger image downloaded from the Internet with low tails, and its head face to the ground,
but the model originally has straight tail in the air, and facing to the front.
The output is a model not only textured by the casual image, but also fit the shape of the tiger image.}
\label{fig:overview}
\end{figure*}

Hence,
in this paper, we propose a novel texturing and deforming method for mapping the pattern and shape of a 2D casual image to a 3D mesh model at the same time 
based on an alternating least-square approach as shown in \figname~\ref{fig:overview}. 
Through a
photogrammetric method, we project the target model onto the source image according to the estimated 
cameras models, which are computed by using Tzur~and~Tal's method~\cite{Tzur:2009}. Then, the target model is deformed according to the shape of the source image using a surface-based deformation method while minimizing the image distortion simultaneously.
The above processes are performed iteratively until convergence.
Hence, our method can achieve texture mapping, shape deformation, and detail-preserving at once, and can obtain more reasonable texture mapped results than traditional methods, since not only the source image is warped to map the target model, but the target model is also deformed to fit the shape of the source image.



The rest of this paper is organized as follows. \secname~\ref{sec:related} describes the most related work of both of texture mapping and shape deformation. We then present our system overview in \secname~\ref{sec:overview} and introduce our main algorithm in \secname~\ref{sec:algo}. \secname~\ref{sec:result} demonstrates some experimental results and provides some discussions. The
conclusion and future work are given in \secname~\ref{sec:conc}.

\section{Related Work}
\label{sec:related}

\textbf{Texture Mapping}
is a very important task for recent computer graphics applications.
There are many approaches to texture 3D meshes with 2D images.
Some of them
generate low-distortion planar parameterization from a mesh surface by dividing the surface into a collection of small patches called atlas that can be unfolded with little 
distortions \cite{Levy:2002,Desbrun:2002,Zhang:2005}.
Another kind of approaches is called constrained parameterization 
\cite{Levy:2001,Kraevoy:2003,Zhou:2005,Tai:2008,Lee:2008}. 
In those methods,
the user needs to specify some one-to-one correspondences between the mesh surface and the image.
Since 
the 3D mesh
and the 2D image may be inconsistent, the results almost have some visual distortions.

Recently, since there are a huge amount of high quality casual images on the Internet, people are starting to think about to use these casual images as the textures.
FlexiStickers~\cite{Tzur:2009} takes the advantages of the constrained parameterization 
with a photogrammetric method to achieve it. 
It solves the problem by figuring the local camera positions for each vertex of the target model. However, the textured models sometimes look strange because of the 
distortion of the images.

For the above texture mapping approaches, although they all tried to minimize the image distortion mathematically, due to the non-similarity between the 2D source image and the 3D target mesh model, the distortion may still be too large.
Hence, if the target model could be deformed a little bit to fit the shape of the source image, it will make the image distortion as much more small as possible.

\textbf{Shape deformation}
is one of the most useful techniques about manipulating and editing the geometry. Recent works about surface-based deformation include multi-resolution mesh editing \cite{Zorin:1997,Kobbelt:1998}, Laplacian or gradient surface editing \cite{Sorkine:2004, Yu:2004, Lipman:2005, Botsch:2008}, and coupled prisms \cite{Botsch:2006}. They all tended to preserve the surface details when directly deforming the target model with some constraints.

The goal of deformation in our system is to manipulate the shape by the reference image while maintaining the surface details of the target model. Our method relies on the linear rotation-invariant (LRI) coordinates~\cite{Lipman:2005}. Inspired by the projection constraint described in Huang~\etal's work~\cite{Haung:2006}, we take the image camera information into the deformation process in order to get the similar projection with the reference image. LRI defined the \textit{discrete form} at each vertex, which records the relative position depends on its one-ring neighborhood. Based on the discrete form, each vertex has its own coordinate frame called local frame. Given the desired location of the features, the restructuring procedure is solving two least-square systems for the local frames and the real position of each vertex. LRI is one of the implicit way to deform the mesh. Because the local frame encodes the orientation of the vertex, the deformation result is rotation-invariant, and the restructuring process forms as a sparse linear problem.

\section{System Overview}
\label{sec:overview}

\begin{figure}
\includegraphics[width=\linewidth]{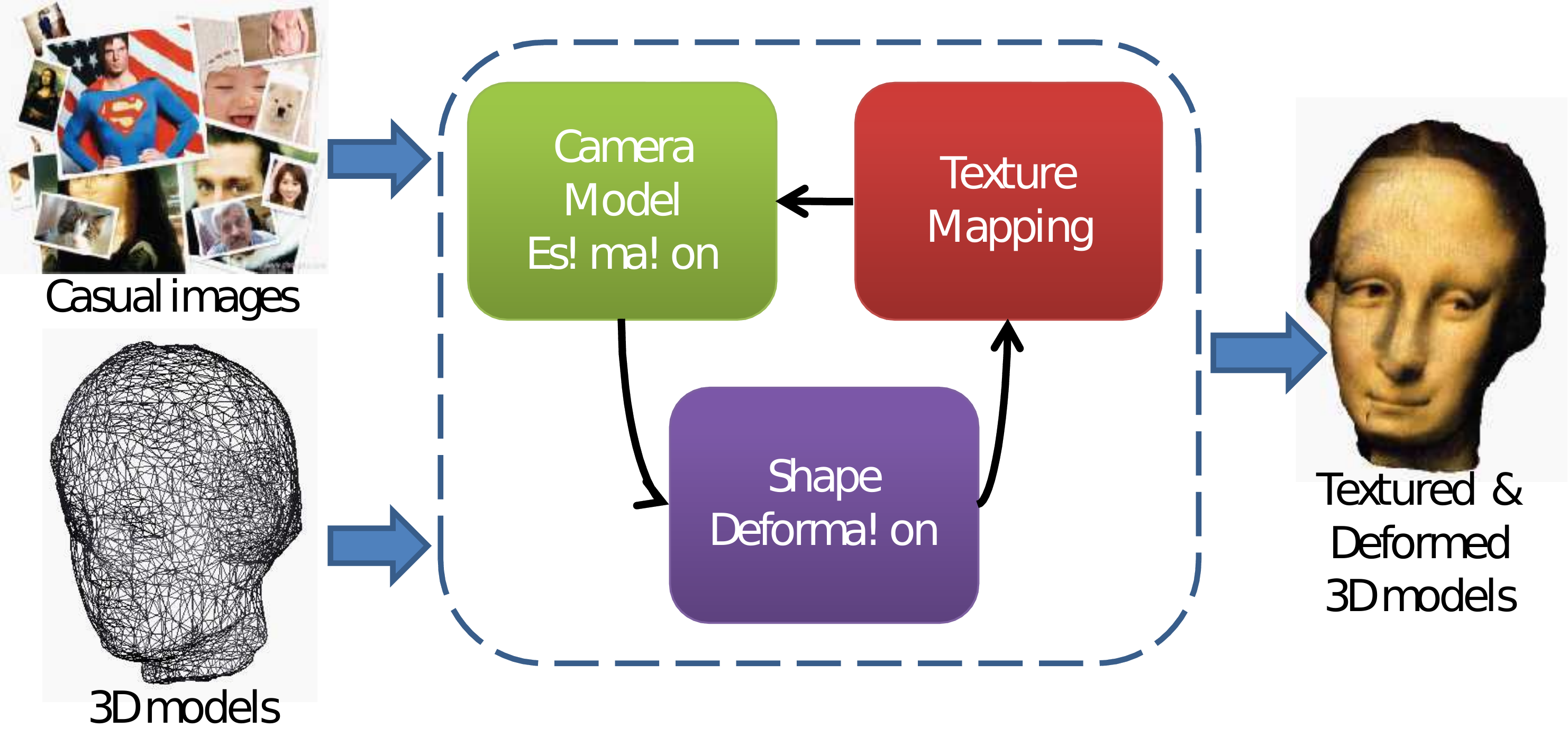}
\caption{The system flowchart of our system. 
}
\label{fig:flow}
\end{figure}

In this paper, we present a novel texturing and deforming approach for mapping the pattern and shape of a casual image to a 3D mesh model at the same time.
This approach achieves texturing, shape manipulation and detail-preserving at once,
and can obtain more visual pleasing and more reasonable texture-mapped results.

\figname~\ref{fig:flow} shows the system flowchart of our system.
Our 
texturing and deforming
approach takes the following three components as the input as several similar previous methods:
\begin{enumerate}
\item A \textbf{casual image}: an image comes from the Internet photo database, which is used as the source texture image.
\item A \textbf{3D model}: a basic template 3D model as the target mesh model.
\item Some \textbf{corresponding feature points}: a set of user specified corresponding points both on the source casual image and the surface of the target 3D model, which will be used for guiding both of the texture mapping and the shape deformation.
\end{enumerate}
Then, 
through
an alternating least-square approach that solves the shape deformation and texture mapping algorithms iteratively and alternatively while minimizing the distortion of the casual image and preserving the surface details of the template 3D model simultaneously.

Inside the system,
there are three major stages of our approach as the follows:
\begin{enumerate}
\item \textbf{Camera Model Estimation}: to estimate a coarse global camera projection model by using 
    Tzur~and~Tal's approach~\cite{Tzur:2009}.
\item \textbf{Shape Deformation}: using the estimated global camera projection model to form the constraints to deform the target 3D model.
\item \textbf{Texture Mapping}: given the updated geometry, each vertex will be textured with the casual image and a new local camera projection model will also be computed.
\end{enumerate}
The above three major processes are performed iteratively and alternatively until convergence.
The result of our texturing and deforming approach is not only a reasonable texture mapped 3D model, but also can obtain the fitted shape 
according to
the casual image.
%
%
\section{Texturing and Deforming Meshes}
\label{sec:algo}

In this section, we elaborate on the details of the entire alternating least-square approach.
In the beginning, we first introduce how we use the same set of constraints (corresponding feature points) for both of the texture mapping and the shape deformation.
Then, the key idea of our approach and the objective function used for optimization are explained.
At last, the alternating minimization strategy to solve the optimization problem is described in details.

Given a casual image $I$ and a 3D template mesh model $M$,
a set
of corresponding feature points $(\lbrace p_i \in I \rbrace_{i=1}^{P} , \lbrace v_i \in M \rbrace_{i=1}^{P})$ is provided by the users, where $P$ is the number of the corresponding features points.
The corresponding feature point set will not only be used for mapping the image onto the mesh, but also for deforming the mesh to fit the image.

\subsection{Objective Function}
\label{sec:obj_overview}

The objective function of our texturing and deforming approach is defined as:
\begin{equation}
E=(1-\alpha)\sum_{i=1}^{N}E_l(v_i)+\alpha\sum_{i=1}^{N}\sum_{j=1}^{P}E_p(v_i, v_j).
\label{eq:obj}
\end{equation}
where $0\leq\alpha\leq 1$ is a weight for controlling the two terms, $N$ is the number of vertices of the mesh model $M$.
The first term in the objective function is used for 
preserving the surface details of the mesh model $M$,
while the second term is used for mapping the texture image $I$ on to the surface of the mesh model $M$.

We will explain each term of this objective function in the following subsections.
Note that adjusting $\alpha$ will influence the importance between the texture mapping and shape deformation phases.
The procedure which comes to the threshold of convergence depends on the weighted error function.
The system will only be operated for texture mapping if we set $\alpha = 1$.

\subsection{Texture Mapping Phase}
\label{sec:texturing}

The 
second term of the objective function \eqname~(\ref{eq:obj})
is used for texture mapping, which
%
is defined as:
\begin{equation}
E_p(v_i, v_j)=\frac{1}{\varepsilon + D(v_i,v_j)^{\beta}}\Vert {\textbf{M}_i}{v_j}+c_i-p_j \Vert^{2},
\label{eq:deo_dist}
\end{equation}
where 
$D(v_i,v_j)$
is the geodesic distance between the two vertices $v_i$ and $v_j$ on the mesh $M$,
$v_j$ and $p_j$
are the corresponding feature points on the mesh $M$
and 
the image $I$, respectively, and
$\beta$ and $\varepsilon$ are small constants as used in \cite{Tzur:2009}.
$\textbf{M}_i\in\mathbf{R}^{2\times 3}$ and $c_i\in\mathbf{R}$ are the world-to-image projection matrix and the translation scalar in the image plane, respectively, which are the unknowns in this term.

Hence, by giving the corresponding points $v_j$ and $p_j$, we tend to calculate the projection matrix $\textbf{M}_i$ and the translation scalar $c_i$ for each vertex $v_i$ on the mesh $M$, so that we can map the image $I$ onto the surface of the mesh model $M$ through the projection matrix $\textbf{M}_i$ and the translation scalar $c_i$.

\subsection{Shape Deformation Phase}
\label{sec:deforming}

The 
texture mapping methods
proposed before only considered the original geometry, even though the object shape in the casual image is quite different from the target 3D model.
If the casual image does not look like the target model, 
the textured model will become strange and the image will be distorted very much.
Hence, we seek to take the shape difference between the geometry and the image into consideration to achieve much more reasonable textured result. 
Therefore,
we introduce a shape deformation term into the objective function \eqname~(\ref{eq:obj}), which is the first term in the equation and is defined
as:
\begin{equation}
E_l(v_{i})=\Vert L(v_{i})-L({v}'_{i})\Vert^{2},
\label{eq:Laplacian}
\end{equation}
where ${v_i}'$ is the updated vertex of the original vertex $v_i$ after the shape deformation, and $L(v_{i})$ 
is the Laplacian coordinate of it as used in similar approaches, like \cite{Lipman:2005}.
Hence, this term seeks to minimize the error caused by deforming the mesh model and could preserve the surface details.

\subsection{Global Camera Estimation}
\label{sec:camera}

For projecting the target model onto the source image, we estimate a coarse global camera model by using
the 
following equation:
\begin{equation}
\argmin_{v_i} =\sum_{j=1}^{P}D(v_i, v_j),
\end{equation}
where $D(v_i, v_j)$ denotes the geodesic distance between a vertex $v_i$ on the mesh $\textit{M}$ and a feature point $v_j$ also on the mesh $\textit{M}$. 
We then find the camera projection of the vertex with the minimum value as our estimated global camera.

\subsection{Alternating Least-Square Optimization}
\label{sec:opt}
Simultaneously optimizing all the unknowns in \eqname~(\ref{eq:obj}) gives rise to an expensive nonlinear least-square problem. However, an initial inaccurate solution to this nonlinear problem can be refined until convergence by iteratively alternating two simpler and more efficient least-square steps, which are respectively responsible for improving the estimation of the vertex positions and the local camera projection models.

The first step of minimizing the objective function is to estimate a global camera model that project the model onto the image. Then, using the estimated global camera model to calculate the vertices' positions $\textit{V}'=\lbrace {v}'_{1},...,{v}'_{N}\rbrace $ after the shape deformation.
The final step does the opposite, that fixes the vertices' positions and optimize the projection matrix $\textbf{M}_{i}$ and the translation scalar $c_{i}$ instead.
The above steps are operated iteratively and alternatively until convergence.

\section{Experiential Results}
\label{sec:result}

\subsection{Implementation}
\begin{figure}[h]
\begin{center}
\includegraphics[width=.8\linewidth]{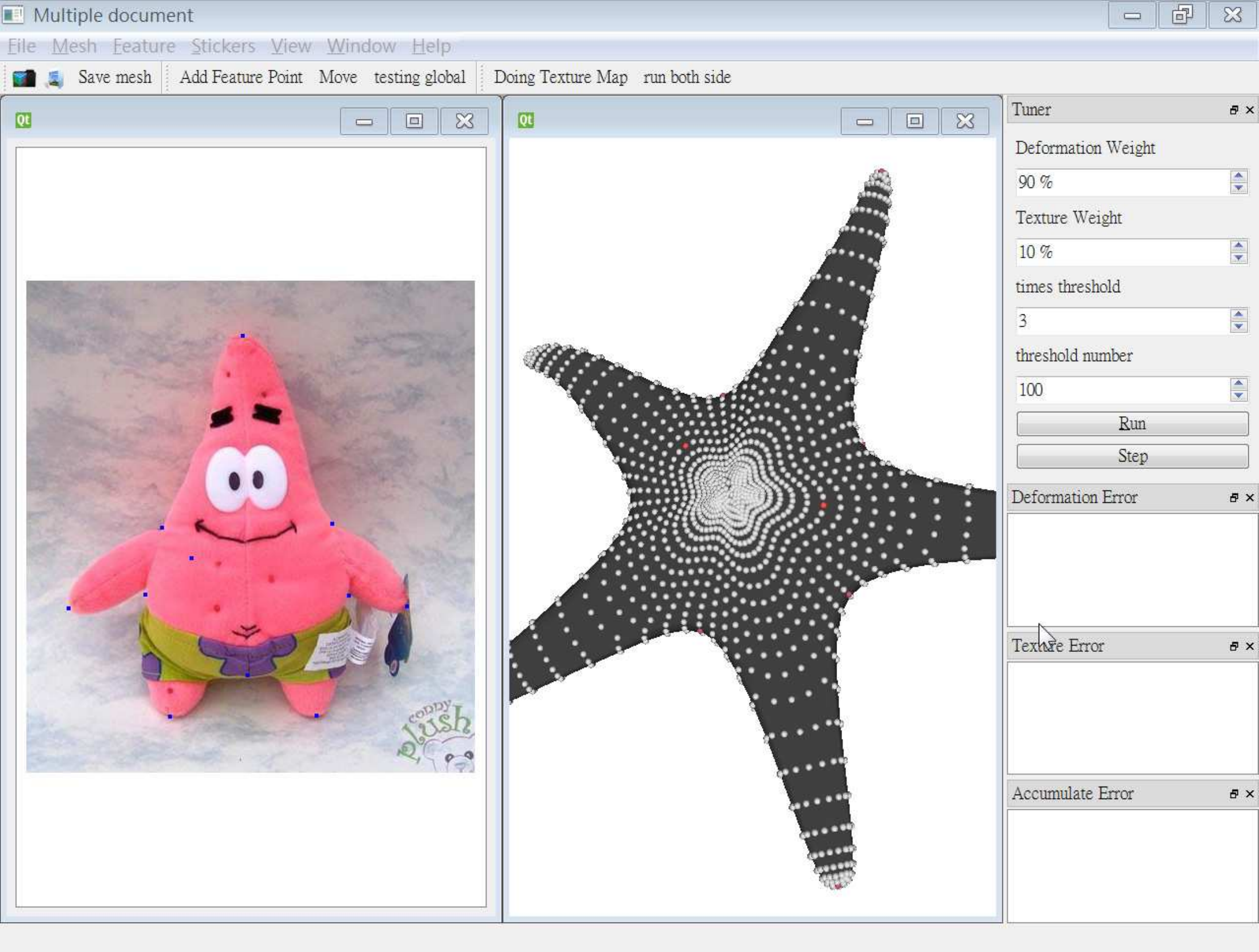}
\caption{A screen shot of the user interface.
}
\label{fig:application}
\end{center}
\end{figure}

Our method was
implemented in C++ with CGAL library and tested on a desktop PC with an Intel Core 2 Q9400 Quad 2.66GHz CPU with 4GB RAM.
The core computation part is solving the sparse linear system, and the \textit{TAUCS} solver is used.
In order to let user to specify the corresponding features between the 2D image and the 3D model, we developed an interactive system as shown in \figname~\ref{fig:application}, which is similar with other texture mapping systems.
This feature-specification user-interface is designed by using QT and QGLviewer library.

\begin{table}[tb]
   	\caption{
        Test data and performance statistics. 
        From left to right: Number of vertices, number of features, number of iterations to converge, and the computation time of each iteration. 
        }
    \label{tab:statistics}
\centering
\begin{tabular}{ c | c c c c}
    \hline
    Model & Vertex\# & Feature\# & Iteration\# & Time(sec.) \\ \hline 
    lion & 5,000 & 86 & 12 & 33.77 \\ \hline 
    halfman & 3,243 & 81 & 4 & 15.79 \\ \hline 
    bottle & 1,441 & 40 & 3 & 12.52 \\ \hline 
    woman & 4,865 & 70 & 4 & 15.24 \\ \hline 
\end{tabular}
\end{table}

\subsection{Result}
\begin{figure*}[htbp]
\centering
\subfigure[]{\includegraphics[height=4.5cm]{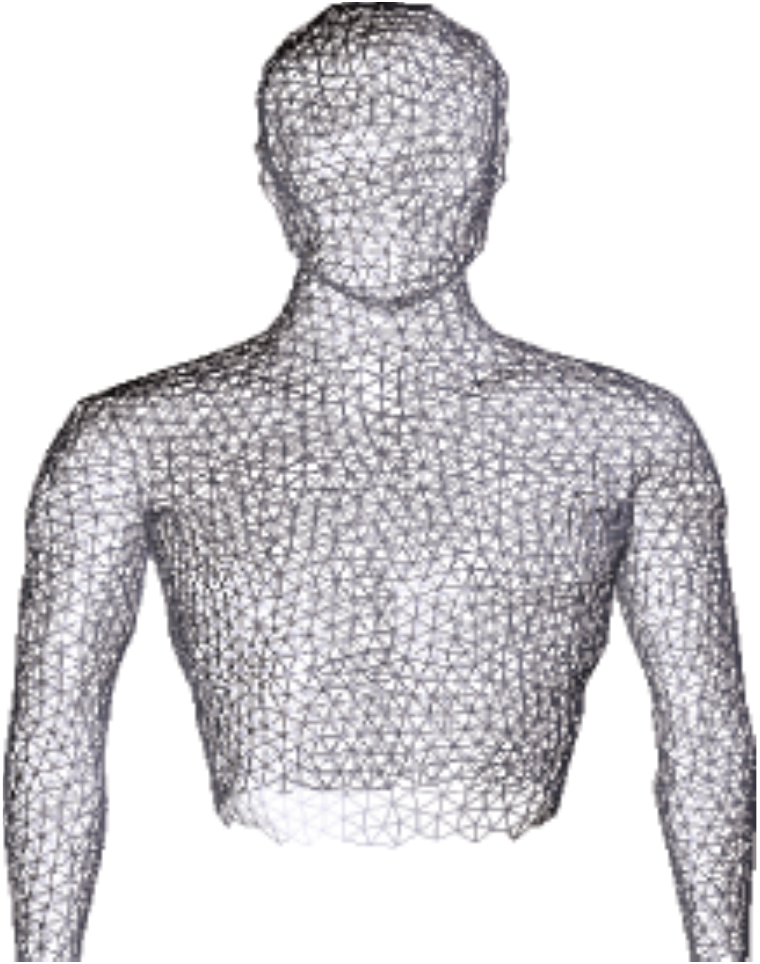}}
\subfigure[]{\includegraphics[height=4.5cm]{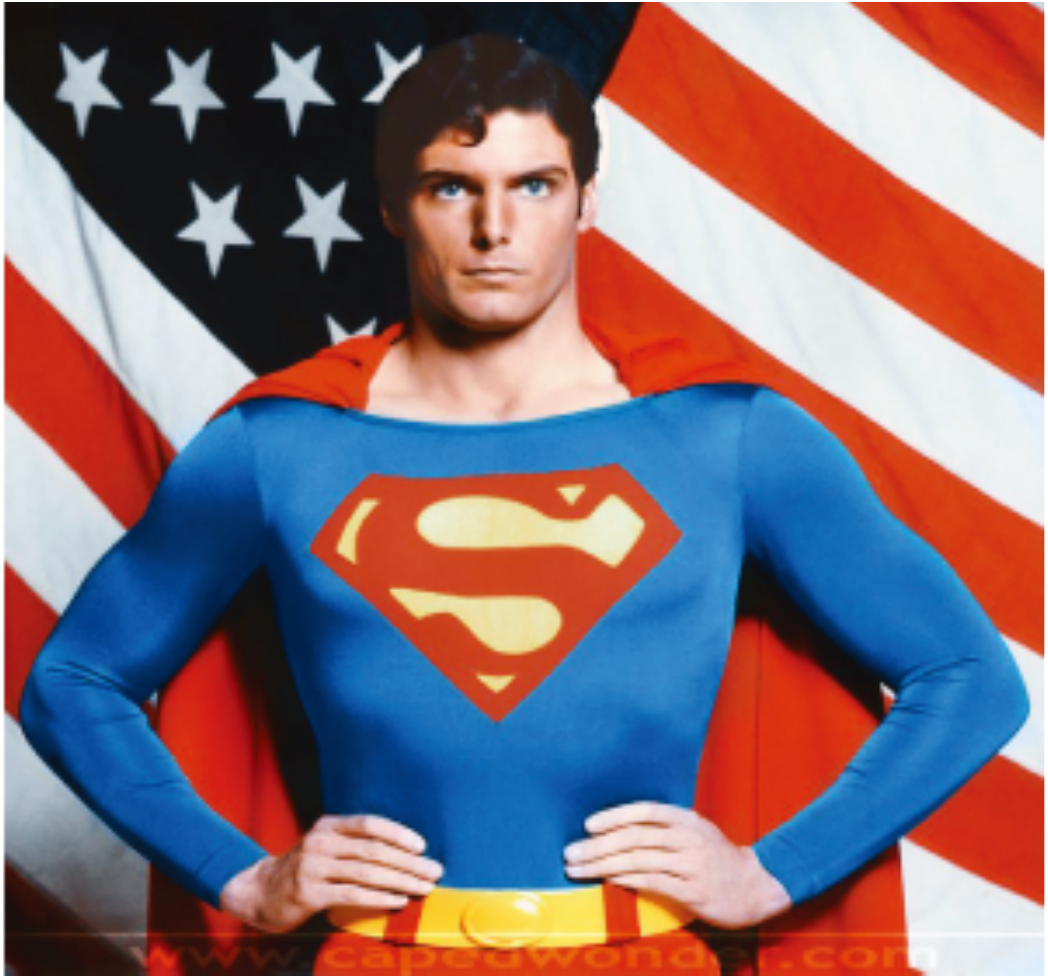}}
\subfigure[]{\includegraphics[height=4.5cm]{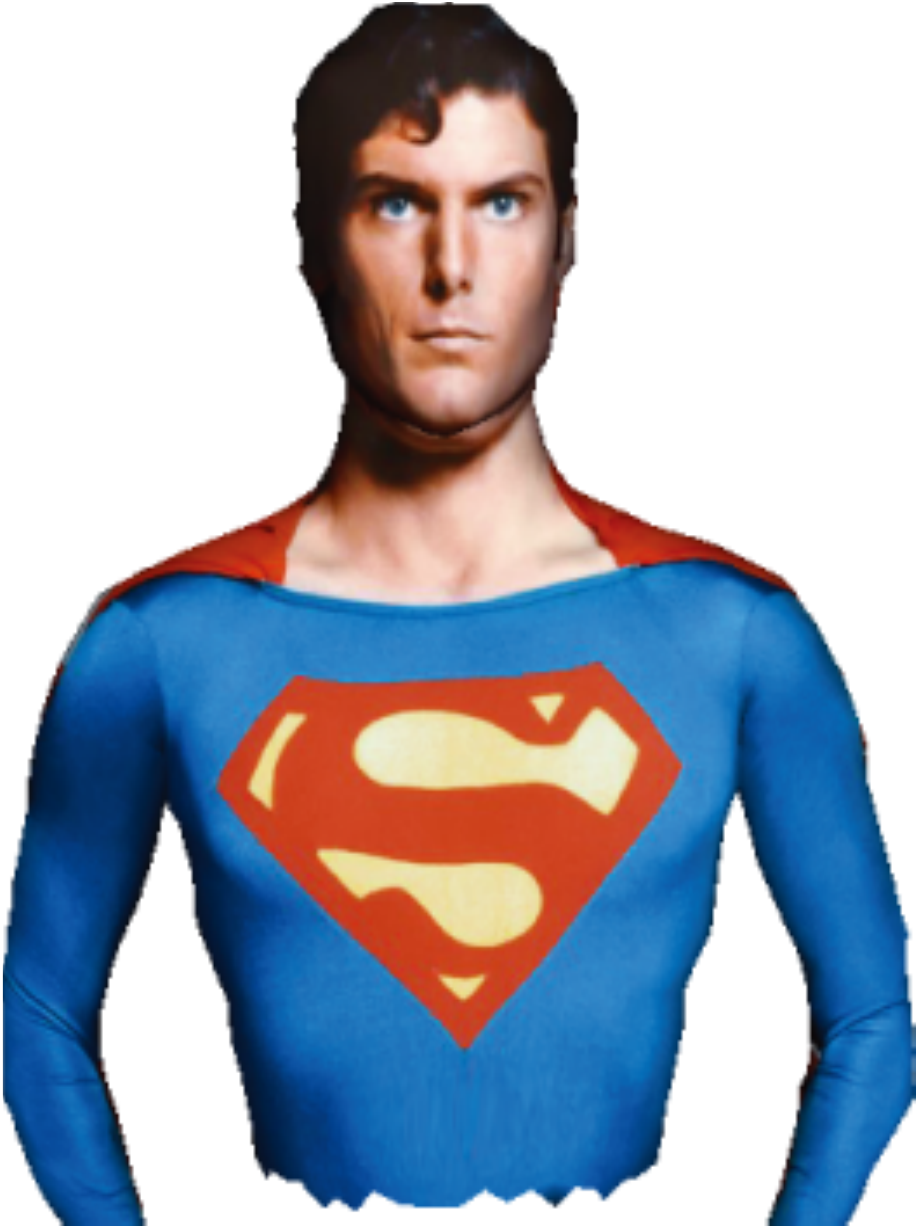}}
\subfigure[]{\includegraphics[height=4.5cm]{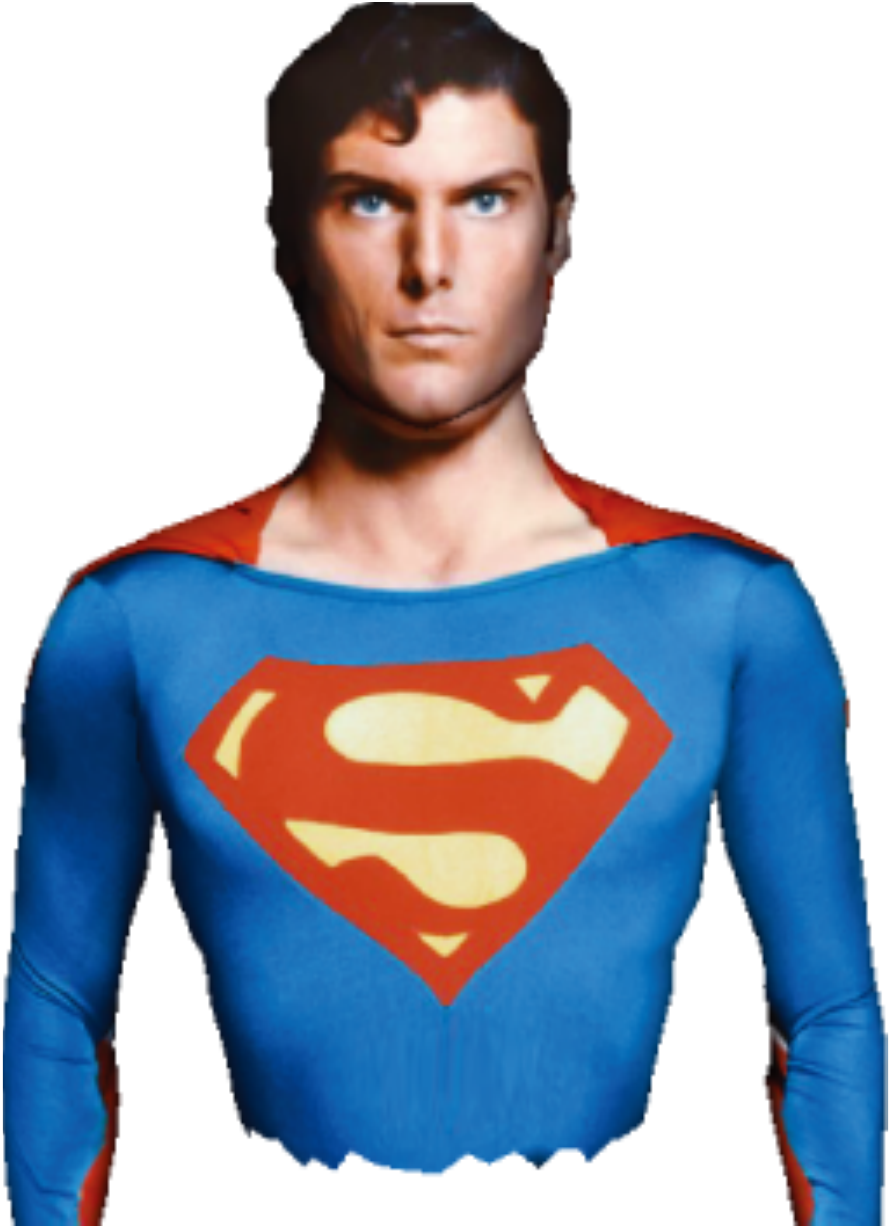}}
\caption{Texturing the half man model using a superman image downloaded form the Internet. (a) The input triangle mesh model. (b) The input image. (c) The texturing (and deforming) result using our algorithm with 86 constraints (feature points). (d) The texturing result of \cite{Tzur:2009}. In this example, our method deformed the elbow on the target model respect to the bend elbow in the source image, and leads to more reasonable textured result at the cloth wrinkle. 
}
\label{fig:superman_compare}
\end{figure*}

In our experiment, the system takes five iterations in average to get the error converged.
The execution time depends on each case as shown in \tabname~\ref{tab:statistics}.
To compute the geodesic distances between the constraints and each vertex takes half of the execution time, and it is directly proportional to the number of the constraints (feature points).
%
\figname~\ref{fig:superman_compare} shows a comparison with our result and the traditional texture mapping result. The pure texture mapping result is not good enough for human visual sense, but
we can improve it by bending the elbow.
\figname~\ref{fig:always_case} shows a simple example by using a simple cylinder bottle to fit to a cup image. Since our method can deform the bottle a little bit to fit the cup shape, the textured and deformed result is better than traditional texturing one.


\begin{figure*}[htbp]
\centering
\subfigure[]{\includegraphics[height=6cm]{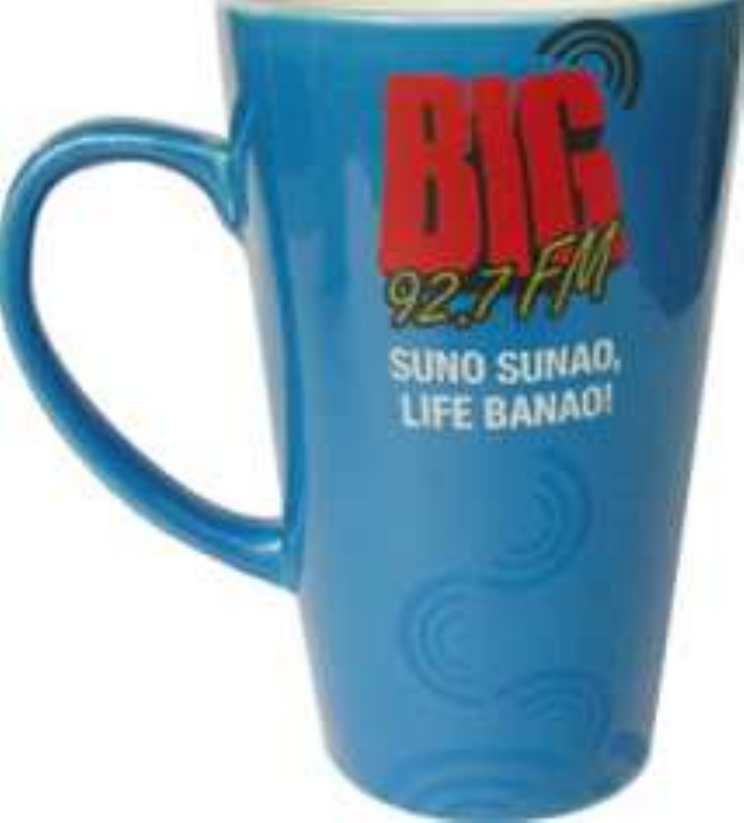}}
\hspace{1cm}
\subfigure[]{\includegraphics[height=6cm]{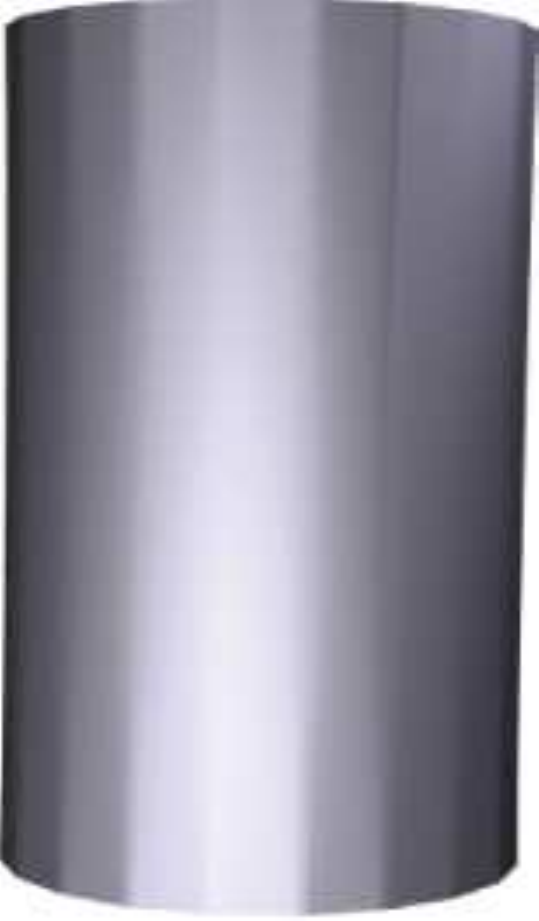}}
\hspace{1cm}
\subfigure[]{\includegraphics[height=6cm]{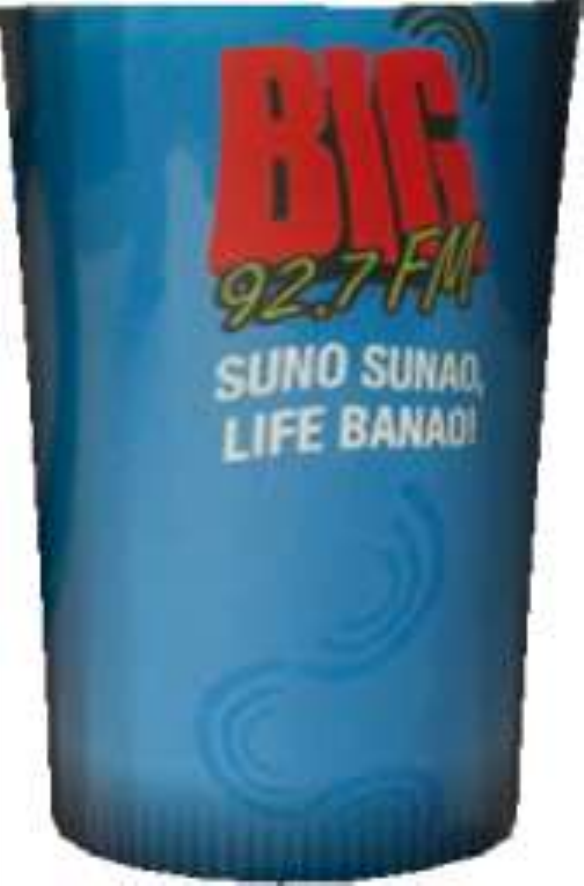}}
\caption{(a) The input cup image. (b) The input simple cylinder bottle model. (c) The textured (and deformed) result.}
\label{fig:always_case}
\end{figure*}

\begin{figure*}[htbp]
\centering
\subfigure[]{\includegraphics[height=6cm]{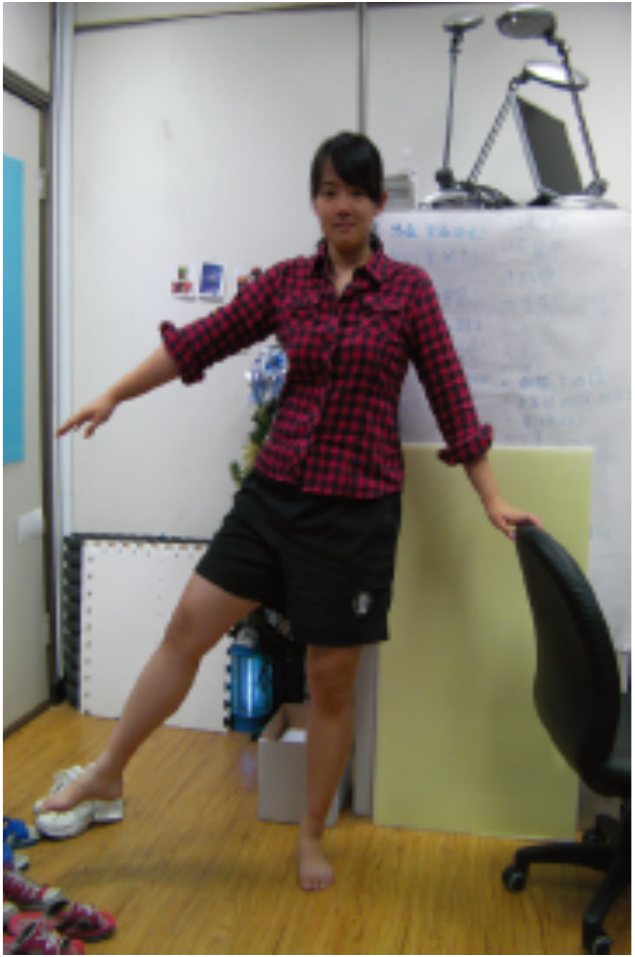}}
\hspace{1cm}
\subfigure[]{\includegraphics[height=6cm]{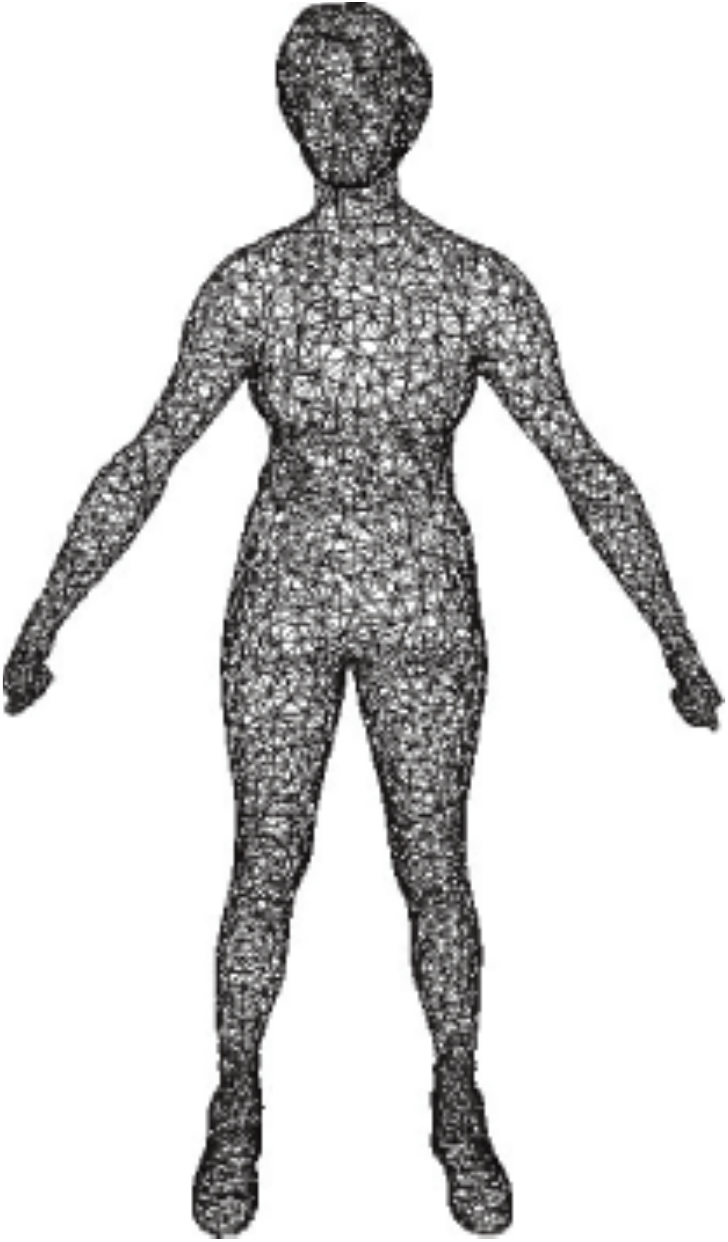}}
\hspace{1cm}
\subfigure[]{\includegraphics[height=6cm]{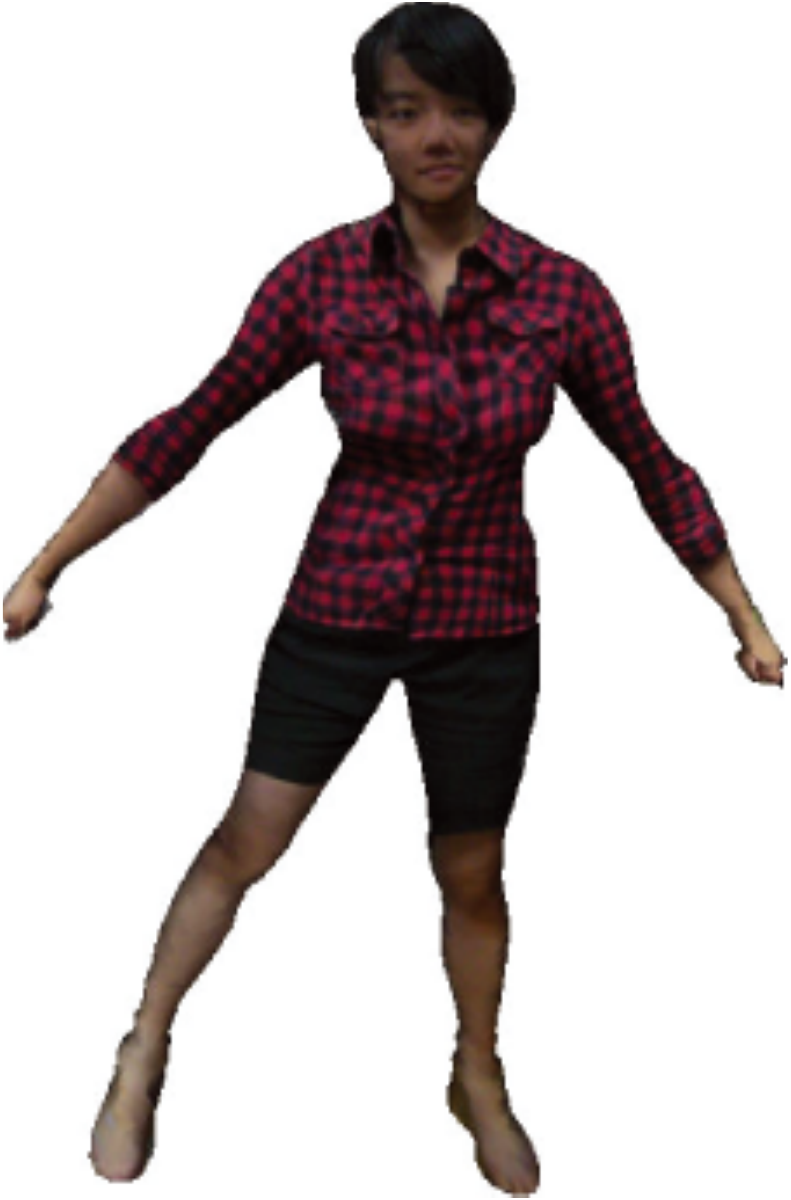}}
\caption{(a) The input woman image. (b) The input simple woman model. (c) The textured (and deformed) result.}
\label{fig:always_case}
\end{figure*}

\subsection{Limitation}
When the image object in the casual image has a quite different shape with the target triangle mesh model,
due to the instinct of the surface-based deformation algorithm, the deformation of the result model still may not be as good as expected.
At the same time, enforcing the deformation may require the user to 
specify many feature points.

\section{Conclusion}
\label{sec:conc}

In this paper, we present a novel framework not only for texturing the casual images onto the 3D models, but also transferring the shape of the image object to deform the models.
Our texturing and deforming method
is based on an alternating least-square approach to optimize the texture mapping and shape deformation problems iteratively.
Since it takes account for the shape of image object additionally, it achieves texture mapping, shape deformation, and detail-preserving at the same time, and can obtain more reasonable texture-mapped results than traditional methods.

In the meanwhile, this framework is naturally composed of two separate algorithms, which are photogrammetric texture mapping and surface-based deformation.
Due to this instinct, it provides great flexibility that everyone can improve the result and accelerate the algorithms in the near future by replacing the texture mapping algorithm or the surface-based deformation one.
In the future, we would like to solve the geodesic distance using GPU, because it is the most time-consuming part in our current system. 

\bibliographystyle{eg-alpha}
\bibliography{pg10-TDM}

\end{document}